\documentstyle[aps,prl,epsf,multicol]{revtex}
\epsfclipon

\draft
\input{psfig}

\begin{document}

\twocolumn[\hsize\textwidth\columnwidth\hsize\csname@twocolumnfalse\endcsname
\title{Vortex Glass and Vortex Liquid in Oscillatory Media}

\author{Carolina Brito $^{1,2}$, Igor  S. Aranson $^3$, and Hugues  Chat\'e $^2$ }
\address{$^1$ Instituto de F\'{\i}sica,
Universidade Federal do Rio Grande do Sul, 91501-970 Porto Alegre, Brazil\\
$^2$ CEA --- Service de Physique de l'Etat Condens\'e,
Centre d'Etudes de Saclay, 91191 Gif-sur-Yvette, France\\
$^3$  Argonne National Laboratory,
9700 South  Cass Avenue, Argonne, IL 60439, USA}
\date{\today}

\maketitle

\begin{abstract}
We study the disordered,
multi-spiral solutions of two-dimensional
homogeneous oscillatory media
for parameter values at which the single spiral/vortex solution
is fully stable.
In the framework of the complex Ginzburg-Landau (CGLE) equation, we show
that these states, heretofore believed to be static, actually evolve
on ultra-slow timescales. This is achieved via a
reduction of the CGLE to the evolution of the sole vortex
position and phase coordinates. This
true defect-mediated turbulence occurs in two distinct phases, a
vortex liquid characterized by normal diffusion of individual
spirals, and a slowly relaxing, intermittent, ``vortex glass''.
\end{abstract}

\vskip1pc]

Spiral waves  are ubiquitous in oscillatory and  excitable
two-dimensional active media \cite{SPIRAL}.
Their cores are robust wave sources which
determine the oscillating frequency of the entire system and may
dominate the surrounding dynamics. Spirals (often called vortices)
may spontaneously appear and annihilate in a typical manifestation
of spatiotemporal chaos. When the single spiral solution
is stable, one easily observes locked, quasi-frozen, multi-spiral
 disordered structures whose glassy character has been
suggested \cite{huber,huberott,chate}. Despite being found in
most models of excitable or oscillatory media as well as in
experiments (Belousov-Zhabotinskii reaction \cite{flessel},
surface growth \cite{abdv}, and others) surprisingly very little
is known about the properties of these disordered states and
the transitions leading to their formation.

In this Letter we investigate, in the framework of the
complex Ginzburg-Landau equation (CGLE),
the general issue of these multi-spiral
solutions and show that they actually evolve
on ultra-slow timescales.
This is achieved thanks both to long numerical simulations of the CGLE
and to a quantitatively correct
reduction of the dynamics to the evolution of the sole vortex
position and phase coordinates. This
true defect-mediated turbulence occurs in two distinct phases, a
vortex liquid characterized by normal diffusion of individual
spirals, and a slowly relaxing, intermittent, ``vortex glass''.

The CGLE describes most properties of generic oscillatory media,
at least at a qualitative level, even if one is not
in the vicinity of a supercritical long-wavelength
Hopf bifurcation, where it can be systematically
derived (for reviews, see \cite{cross,arkr}). Under
appropriate scaling of the physical variables, it takes the
universal form
\begin{equation}
\partial_t A=A-(1+ic)|A|^2A+(1+ib) \Delta A,
\label{cgle}
\end{equation}
where $A$ is a complex amplitude, $b$ and $c$ are real parameters
characterizing relative dispersion and nonlinear frequency shift,
and $\Delta$ is the Laplace operator.
Intensive studies conducted over the last ten years
revealed a wide variety of striking dynamical phenomena  in one,
two, and three space dimensions, many of which were also observed in
various experimental contexts, sometimes up to a quantitative
agreement with CGLE predictions.

A distinctive feature of the two-dimensional CGLE is the existence
of nontrivial sources of spiral waves (vortices) which determine
the oscillating frequency of the entire system \cite{kura}. The
single-spiral solution  for a vortex centered at ${\bf r}_0$ reads
as: \cite{PSH,akw}:
\begin{equation}
A_{\rm s}({\bf r},t)= F(r) \exp [ i(m \theta + \psi(r) + \omega t+\phi)],
\label{spir}
\end{equation}
where $r=|{\bf r}-{\bf r}_0|$, $\theta$ is the polar angle
measured from the vortex core, $\phi$ is an arbitrary phase,
and $m=\pm1$ is the topological charge.
Far away from the core, the solution approaches a  plane wave with
$\psi(r)\simeq kr$, where the asymptotic wavenumber $k$ is related
to the rotation frequency as $\omega=-c-(b-c)k^2$.
The dependence of $k$  on
$b$ and $c$ is known analytically for $|b-c|  \ll 1$  and $|b-c| \gg 1$,
e.g. $k\simeq - c^{-1} \exp(-\pi/|2c|)$ for $b=0$ and $|c|\ll 1$ \cite{PSH}.
In the parameter region where the single-spiral solution is stable
and  $c \ne b$, the interaction
between two well-separated spirals falls off exponentially
\cite{akw,bick,pn}.
The screening was attributed to the
shock lines where the waves emitted by the cores collide. For
relatively small $b,c$ the interaction is monotonic, and the spirals
exhibit asymptotic repulsion irrespectively of their charge
(``monotonic range'')
\cite{akw,bick}. In contrast, for larger $b,c$ satisfying the condition
$(c-b)/(1+bc)>c^*\approx0.845$ the velocity vs distance dependence
is modulated  (``oscillatory range''),
and the spirals become keen to form a variety of long-living (but unstable)
bound states \cite{akw,akw1}.

While the above interactions cannot account for the
strongly chaotic regimes where many defects are spontaneously
generated and undergo violent motion, they are expected to
play a leading role in
the occurrence of the quasi-static structures of large and small
spirals surrounded by a complex network of shocks commonly
observed in the large region of the $(b,c)$ parameter space
where the single spiral solution is stable
\cite{huber,huberott,chate}.
Rather little is known about these cellular
structures except that they seem to be frozen on usual observation
timescales. Our results lead to a dramatically  different
picture.

We conducted detailed numerical studies of the two-dimensional
CGLE restricted  to the case $b=0$ and $c>0$ (for the regimes of
interest scaling relations apply in parameter space \cite{arkr}).
The integration domain with periodic boundary conditions was
typically of area $S=512\times512$ and the integration time about
$10^7$. Various integration schemes were used and all results
checked against variation of the numerical resolution. We
monitored the positions ${\bf r}_j(t)$ of  all $N$ spirals. We
measured their instant ``activity'' $T
=S^{-1}\int_S d{ \bf  r}   \left| \partial_t |A| \right|$ and
---when applicable--- the spiral diffusion coefficient $D=
\frac{1}{Nt}   \sum_{j=1}^N \left\langle | {\bf r}_j (t)-{\bf
r}_j(0) | ^2 \right \rangle$.
Since, for well-separated spirals, $|A|$ is constant everywhere except
near the cores, $T$ is  related to the velocity of spirals.

\begin{figure}
\centerline{ \epsfxsize=8.6cm \epsffile{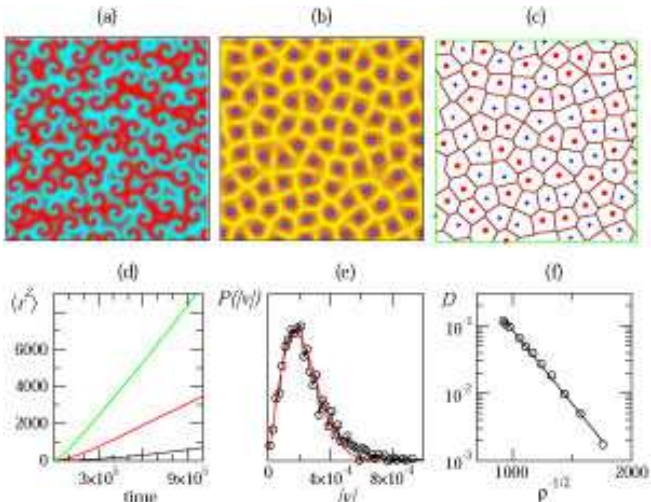}} 
\caption{Vortex
liquid in the monotonic range, for the full CGLE (a,b,d,e) and for
Eqs.(\ref{sys1}) (c,f). (a,b) snapshots of ${\rm Re}(A)$ and $|A|$
for a typical solution of the CGLE at $c=0.6$ with $N=82$ spirals
(domain of size $512\times 512$ with periodic boundary
conditions). (c) Snapshot of core positions and shock lines for a
typical solution of Eqs.(\protect \ref{sys1}) for the same
parameters as in (a,b) (d) Mean square displacement $\langle r^2
\rangle$ of vortices vs time obtained from solutions of the CGLE
for $c=0.7$, $L=256$, and $N=28,22,18$ vortices from top to
bottom. (e) Distribution function of the instantaneous spiral core
velocity (arbitrary units) with fit by a Maxwellian from a
simulation of the full CGLE at $c=0.7$ with 118 vortices in a
$512\times 512$ domain. (f) Variation of $D$, the diffusion
constant of vortices, with
 $1/\sqrt{\rho}$ for $c=0.3$  (from
solutions of  Eqs.(\protect \ref{sys1})).
\label{fig1}
}
\end{figure}

In the monotonic range $c<c^*$ (Fig.~\ref{fig1}),
after a short transient during which
the initial number of defects may decrease,
the number of spirals remains constant and
$T$ fluctuates around a well-defined mean value.
The typical vortex velocity is very small
(of the order of $10^{-4}$ for $c=0.7$), and the velocity
distribution is approximately Maxwellian (Fig.~\ref{fig1}d).
In  large enough systems,
the spiral cores perform normal diffusive motion
(Fig.~\ref{fig1}d). This {\it vortex liquid}
can be characterized by a viscosity  $\nu\sim D^{-1}$.
Since all scales, already very large near $c^*$,
diverge exponentially when $c\to0$,
an extensive investigation of the dependence of $D$ on $c$ and on
the vortex density $\rho\equiv N/S$ is beyond reach of present-day
computers. Nevertheless, our data indicate that $D$ increases with $\rho$.

\begin{figure}
\centerline{
\epsfxsize=8.6cm
\epsffile{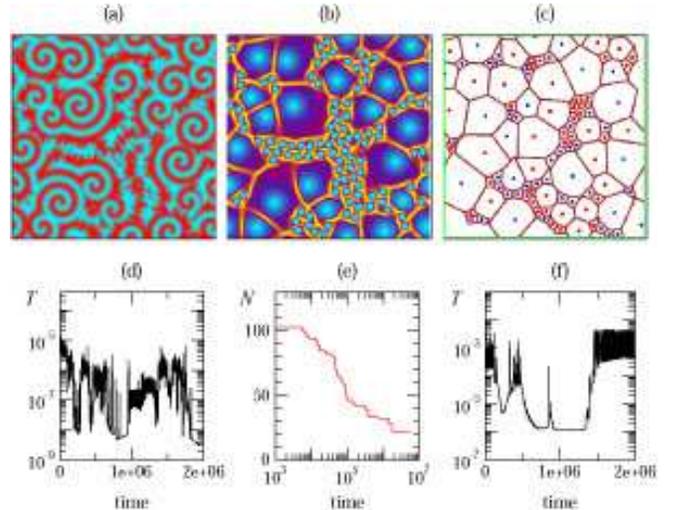}}
\caption{Results in the oscillatory range ($c=1.2$), obtained on the
full CGLE (a,b,d,e) and from the reduced equations (c,f).
(a,b) snapshots of ${\rm Re}(A)$ and $|A|$ for a typical solution of
the CGLE with $N=124$ spirals
(domain of size $256\times 256$ with periodic boundary conditions).
(c) Solution of Eqs.(\protect \ref{velxy}) for the same parameters
as in (a,b).
(d,e): time series of the activity $T$ and of the number of vortices
$N$ for the evolution of the CGLE in a domain of size $L=256$
starting from random initial conditions.
(f): same as (d), but from simulations of Eqs.(\protect \ref{velxy})
with $N=100$ vortices
(in this case $T\equiv N^{-1} \sum_j |{\bf v}_j| $)
\label{fig2}
}
\end{figure}

In contrast, in the oscillatory range $c>c^*$,
after a very long, possibly infinite, transient
(Fig.~\ref{fig2}e), the population
of spirals  spontaneously segregates into two distinct phases:
large and almost immobile spirals and droplets (clusters) of small
vortices confined between them (Fig.~\ref{fig2}a,b, see also
\cite{huber,huberott,chate}). One can thus define a
``liquid fraction'' of small spirals, whose sizes are typical of the
vortex liquid observed in the monotonic range. When the liquid
fraction is small, the resulting state exhibits slow intermittent
dynamics (bursts of activity separated by long quiescent
intervals, Fig.~\ref{fig2}d) reminiscent of glassy dynamics,
whereas for higher fractions the activity $T$ fluctuates around
some mean value and the ``liquid'' vortices can be shown to
perform normal diffusion in the space surrounding the big spirals.
This picture is reminiscent of phase coexistence in first-order
transitions of isolated equilibrium systems  \cite{ls}. In
this respect, our results indicate that one can indeed distinguish
a ``vortex glass'' phase. Unfortunately, because of the ever slower
timescales over which the spatial structure of the system evolves,
a precise characterization of this phase using
simulations of the CGLE is currently impossible.

To obtain further insight into the problem, we reduced the CGLE to
a set of ordinary differential equations describing the motion  of
the cores ${\bf r}_j(t)$ and the  phases $\phi_j(t)$ of individual
spirals. Only pair interactions are taken into account, making use
of the results of \cite{akw,akw1}. The original partial
differential equation is thus replaced by a set of $3N$
first-order equations governing the core positions ${\bf r}_j(t)$
and the spiral phases $\phi_j(t)$. The phases determine the
configuration of the shock lines, which, in turn, yield the
velocity of the cores. One clear advantage of this approach,
in addition to virtually suppress physical space,
is that it allows to bypass the ultraslow timescale related to the
vanishing value of the wavenumber $k$ as $c\to 0$.

The problem of the interaction between two oppositely-charged
spirals is equivalent to that of one spiral and a straight shock
line, i.e. a half-plane with the no-flux boundary condition
$\partial A/\partial n=0$. For  $c \ll 1 $,   the equations
governing the position and phase of the spiral are
\begin{eqnarray}
\frac{dz}{dt}&=&2 c^2 \pi k (i m-  c B^\prime) \frac{\exp (-2 | ck| X) }{ \sqrt{ \pi |ck| X}} , \nonumber \\
\dot{ \phi}&=& 2 ck^2 \pi \frac{\exp (-2 | ck| X) }{ \sqrt{ \pi
|ck| X}} \label{sys}
\end{eqnarray}
where $z=x+i y$ is the (complex) spiral position,  $dz/dt=v_x+i v_y$,
$\dot{\phi}$ is the phase correction, $m$
is the topological charge, $B^\prime\simeq 0.48$, and $X$ is the
distance from the core (located at $(-X,0)$) to the boundary (the
y-axis is chosen to be on the shock line) \cite{bick}.
Eqs.~(\ref{sys}) imply the asymptotic repulsion of the spiral from
the boundary, in agreement with simulations \cite{akw}. This
prevents spiral annihilation in the vortex liquid.

The  equations in the oscillatory range $c>c^*$ read:
\begin{eqnarray}
&&C_x v_x+ m C_y v_y  =\frac{-k \sqrt{1-k^2} \exp(-p X)} {\delta
\sqrt{2\pi p X }} X^{- \mu}  ,
\label{velxy} \\
\dot{ \phi} &=&\frac{1}{{\rm Im} (C_{10} C_0^*)} {\rm Im}
\left(\frac{-k \sqrt{1-k^2} \exp(-p X)} {\delta \sqrt{2\pi p X }}
X^{- \mu} C_0^* \right) \nonumber
\end{eqnarray}
where the complex  constants $C_{x,y},C_0,C_{10}$ are obtained
numerically and the real parameters $p,\mu,\delta$ are derived
from the linearized CGLE \cite{akw,akw1}. Eqs. (\ref{velxy})
describe stable bound states of spirals moving along the plane
boundary. However, the problem is complicated by the fact that the
position of the shock line (where the waves emitted from the cores
collide)   depends on the relative phase of the spirals. From the
condition that the total spiral phases $\Phi_j=\psi(r) +\phi_j
\approx -|k| r + \phi_j $ are equal, one finds that the distance
to the shock is given by:
\begin{equation}
X=|{\bf r}_k-{\bf r}_j|/2 -(\phi_k-\phi_j)/2|k| \quad
\label{shock}
\end{equation}
Substituting (\ref{shock}) into Eqs. (\ref{velxy}) one
finds that the symmetric bound states are unstable in the oscillatory
range \cite{akw1}. Similar phenomena occurs for
likely-charged spirals.

Eqs. (\ref{sys}), (\ref{velxy}) and condition (\ref{shock}) were
used to investigate the dynamics of many-spiral states. In this
case, one considers each spiral in the local coordinate system
associated with every other spiral and sum up their  contributions
(details will be published elsewhere). Substituting $r \to 2
c|k|r, t  \to 4 \sqrt {2 \pi} c^3 k^2$ and introducing a rescaled
``phase length'' $\zeta=c \phi$, one obtains from Eqs.(\ref{sys}):
\begin{eqnarray}
\frac{dz _j}{dt}  &=&\sum_k (-c B^\prime+im_k ) \frac{z_k
-z_j}{|z_k -z_j|}
\frac{\exp (- X_k) }{ \sqrt{  X_k}}, \nonumber \\
\dot{ \zeta_j}&=&  \sum_k
 \frac{\exp (-X_k) }{2 c \sqrt{  X_k}}
\label{sys1}
\end{eqnarray}
where  $X_k=|z_k-z_j|/2 +\zeta_j-\zeta_k$. A similar rescaling was
performed
in the oscillatory case.

We investigated numerically and analytically the reduced equations
which, in fact, are rather different in the oscillatory range
\cite{akw,akw1} and in the monotonic range \cite{bick}. Overall
quantitative agreement with the full CGLE dynamics has been found,
which constitutes maybe the first success for the ideas underlying
the concept of ``defect-mediated turbulence'' put forward in the
80's \cite{arkr,coullet}.

In particular, we found, in the monotonic range, clear ``liquid''
behavior accompanied by normal diffusion for all $c$ values. Each
spiral core moves chaotically and the velocity distribution is
also very close to a Maxwellian. The reason underlying the chaotic
behavior of this vortex liquid is that Eqs. (\ref{sys1}) do not
obey a variational principle for any $c$ value, due to the
nontrivial form of the pairwise interaction. However, an ensemble
of likely-charged spirals tend to form a stable hexagonal
``Wigner'' crystal due to mutual repulsion. With periodic boundary
conditions (used here), one has an equal mixture of positive and
negative spirals, and it can be proved that a square lattice of
spirals with alternating charges is unstable in a large system
with respect to long-wavelength perturbations. This explains the
short-range crystalline order and local spatial charge separation
observed (Fig.~\ref{fig1}c). The faithfulness of the reduced
equations to the CGLE is also testified by the {\it quantitative}
agreement: found for the value of the diffusion constant $D$. for
$c=0.6$, the domain size $L=512$ and vortex number $N=46$ the
diffusion $D$ obtained from CGLE is   $D \approx 0.0033$  and from
Eqs. (\ref{sys1}) is $D \approx 0.0036$.
The reduced equations do allow for an extensive study of
the asymptotic (i.e. large-size) properties of the vortex liquid.
In particular, we find that
$D \propto \exp(-2c|k|/\sqrt{\rho})$ at fixed $c$ value
(Fig.~\ref{fig1}f).
Taking into account that the typical inter-spiral distance varies
like $1/\sqrt{\rho}$,
this scaling law is in agreement with the expression for the spiral
velocity as given by Eq.(\ref{sys}).

In the oscillatory range also, the reduced equations
faithfully reproduce the phenomenology of symmetry breaking,
intermittent activity at low vortex densities, and
``vortex glass'' formation observed in the CGLE (Fig.~\ref{fig2}c,f).
The reduced equations provide an interesting framework to discuss
the mechanisms leading to intermittent dynamics in
the vortex glass containing liquid droplets.
For a spiral embedded in a large domain of arbitrary shape,
Eqs. (\ref{velxy}) generally do not have a stationary solution
${\bf v}_j=\dot{\phi}_j=0$ for all $j$ (because
the complex constants $C_{x,y} $
and $C_{10,0}$ are typically not proportional). As a result, it is impossible
to satisfy the conditions ${\bf v}_j=0$ and $\dot{\phi}_j=0$
simultaneously. The drift of the spirals changes the phases
and, therefore, the shape of their domains.
Eventually, when the phase difference becomes large enough,
the shock line boundaries rearrange and trigger rapid dynamics
within the liquid droplets.
Since the velocity of the cores is an exponential function of the inter-spiral
separation,
the typical timescale between such rapid rearrangement events
is exponentially large in the spiral size.
Thus, in very large systems, where the domains occupied by spirals may be
comparable with system size, the dynamics will be very slow indeed.

We have found that the transition from vortex liquid
to vortex glass possesses some features of a
first order phase transition.
Additional numerical studies will be needed to precise the structure
of corresponding phase diagram.
Returning to the question of the possibility of ``true''
glassy dynamics \cite{mezard}
(as testified, e.g., by some aging phenomena), several comments are
now in order. Preliminary numerical results of the reduced equations
in very large systems at low vortex density indicate that the intermittent
space-time activity usually reaches a complex but
self-averaging asymptotic regime,
in contrast with spin-glass-type behavior \cite{fh} and in closer agreement with
a structural glass, which can be viewed sometimes as a very viscous fluid.
On the other hand, these regimes,
as far as they can be studied for the full CGLE,
do exhibit a slow, possibly aging-like, decay of the number of vortices
in the system (Fig.~\ref{fig2}e).
This leaves the door open for an actual vortex glass
in this fully-deterministic, noiseless, disorder-free context
--- a challenge to models of glassy behavior in statistical physics.

Our work has shown that the multi-spiral ``frozen'' states of the CGLE
actually evolve on very long timescales. These ultra-slow regimes
are relevant, we believe, to the many related models and experimental
situations. A vortex liquid and a vortex glass can be
distinguished depending on the effective interaction law between
spirals. These ``phases'' are dynamical, and are remarkably well
accounted for, in the case of the CGLE, by a reduction of the full
problem to a finite set of ordinary differential equations
governing the vortex cores. Given that the typical timescales of
vortex motion are several orders of magnitude larger than the
period of the basic oscillations, any experimental observation
will require a very good control of the apparatus over long times.
This work was supported by the U.S.Department of Energy under
contract W-31-109-ENG-38.

\references{

\bibitem{SPIRAL} A. T. Winfree,
{\em When Time Breaks Down}, (Princeton U. P., Princeton, NJ, 1987);
 A. Goldbeter, {\em Biochemical Oscillations and Cellular
Rhythms}, (Cambridge U. P., Cambridge, 1996).

\bibitem{huber}
G. Huber, P. Alstr{\o}m and T. Bohr,  \prl{\bf 69} 2380 (1992)

\bibitem{huberott}
T. Bohr, G.  Huber, and E. Ott,
   Europhys. Letters {\bf 33}, 589 (1996)

\bibitem{chate}
H. Chat\'e  and P. Manneville, Physica A {\bf 224} 348 (1996).

\bibitem{flessel} Q. Ouyang and J.M.~Flesselles, Nature {\bf 379}, 143 (1996)

\bibitem{abdv} M. Hawley, I.D. Raistrick, J. G. Beery, R.J. Houlton, Science {\bf 251}, 1587 (1991);
I.S. Aranson, A. R. Bishop, I. Daruka and V.M. Vinokur, \prl {\bf
80}, 1770 (1998)

\bibitem{cross} M. Cross and P.C. Hohenberg,
\rmp {\bf 65} 851 (1993)

\bibitem{arkr} I.S. Aranson and L. Kramer,  Rev. Mod. Phys. {\bf 74}, 99 (2002)

\bibitem{kura} Y. Kuramoto, {\em Chemical Oscillations, Waves and
Turbulence}, (Springer, Berlin, 1984)

\bibitem{PSH}P.S. Hagan,
SIAM J. Appl. Math. {\bf 42}, 762 (1982)

\bibitem{akw} I.  Aranson, L. Kramer and A. Weber,
 \pre {\bf 47}, 3231 (1993)

\bibitem{bick} V.N. Biktashev,
"Drift of reverberator in active media due
to interaction with boundaries",
{\it Nonlinear Waves II}, eds. A.V.
Gaponov-Grekhov and M.I. Rabinovich, Research Reports in Physics
(Springer, Heidelberg, 1989)

\bibitem{pn} L.M. Pismen and A.A.\ Nepomnyashchy,
    Phys. Rev. A {\bf 44}, 2243 (1991)

\bibitem{akw1} I.  Aranson, L. Kramer and A. Weber,  \pre  {\bf 48}, R9 (1993)

\bibitem{ls} A.J. Bray, Adv. Phys. {\bf 43}, 357 (1994)

\bibitem{coullet}P. Coullet, L. Gil, and J. Lega, \prl {\bf 62}, 1619 (1989)

\bibitem{mezard}
C.A. Angell, Science, {\bf 267}, 1924 (1995); M. M\'ezard, First
Steps in Glass Theory, in {\it More is different}, M. Phuan Ong
and Ravin N. Bhatt eds., Princeton University Press, 2001.

\bibitem{fh} K.H. Fischer and J.A. Hertz, {\it Spin Glasses}
(Cambridge University Press, London, 1991)

}

\end{document}